\documentclass[11pt]{article}
\usepackage{hyperref}
\pdfoutput=1

\begin{document}

\title{A Retrospective on Modulated Wavy Vortex Flow}

\author{Michael Gorman \\
\vspace{6pt} Department of Physics\\University of Houston\\
Houston, TX 77204-5005\\ and \\Harry Swinney\\\vspace{6pt}
Department of Physics, \\University of Texas, Austin, TX 78712}

\maketitle


\begin {abstract}
{A fluid dynamics video of the Modulated Wavy Vortex Flow state of 
Taylor-Couette flow with the outer cylinder fixed is presented. This state 
precedes the transition to turbulence, which is more gradual than that for 
other fluid systems.}

\end{abstract}




\href{http://hdl.handle.net/1813/13744}{A Retrospective on Modulated Wavy Vortes Flow.mpg}

Circular Couette flow with the outer cylinder fixed undergoes three bifurcations 
before the onset of turbulence.  The initial steady, azimuthal flow undergoes a 
bifurcation to Taylor vortex flow in which $N$ vortices form along the axis of rotation.  
At a second critical value of the rotation rate, these flat vortices become unstable 
to the formation of M waves traveling around the annulus.  Our experiments followed 
on the work of Gollub and Swinney, who observed a second characteristic frequency 
before the onset of turbulence.  We found that the second frequency corresponded to 
the independent modulation in the shape of each wavy vortex subject to the condition 
that the phase, $\phi$, between successive modulations equals $K\cdot 2\pi /M$.  We present video 
(recently converted from $16$ mm film and never before presented publically) of three 
MWVF states.  Circular Couette flow is an example of a fluid system which becomes 
turbulent more gradually than most other fluid flows.

%
\end{document}